\documentclass[journal]{IEEEtran}
\usepackage{amsmath,amssymb}
\usepackage{amsfonts}
\usepackage{epsfig,graphicx}
\usepackage{color}

\usepackage{enumerate}

\DeclareMathOperator*{\sgn}{sgn}

\newcommand{\Prob}{\mathrm{P}}

\newcommand{\rme}{\mathrm{e}}

\newcommand{\rmd}{\mathrm{d}}

\newcommand{\setI}{\textbf{I}}

\newcommand{\1}{\mathbf{1}}

\newcommand{\rmf}{\mathrm{f}}
\newcommand{\rmF}{\mathrm{F}}
\newcommand{\rmg}{\mathrm{g}}
\newcommand{\rmm}{\mathrm{m}}

\newtheorem{theorem}{Theorem}[section]
\newtheorem{lemma}[theorem]{Lemma}

\newtheorem{corollary}[theorem]{Corollary}

\begin{document}

\title{On reliable computation by noisy\\ random Boolean formulas }

\author{Alexander~Mozeika  and~David~Saad
\thanks{Alexander Mozeika is currently with the Institute for Mathematical and Molecular Biomedicine, King's College London, Hodgkin Building, London SE1 1UL, United Kingdom. e-mail: alexander.mozeika@kcl.ac.uk; the work had been carried out when he was with the Non-linearity and Complexity Research Group, Aston University.} 
\thanks{David Saad is with the Non-linearity and Complexity Research Group, Aston University, Birmingham, B4 7ET, United Kingdom.}
\thanks{Manuscript received June X, 2012; revised November X, 2014.}}

\markboth{ieee transactions on information theory,~Vol.~X, No.~X, May~2012}%
{Shell \MakeLowercase{\textit{et al.}}: Bare Demo of IEEEtran.cls for Journals}

\maketitle

\begin{abstract}
\boldmath We  study noisy computation in  randomly generated k-ary Boolean formulas. We establish bounds on the noise level above which the results of computation by random formulas are not reliable. This bound is saturated by  formulas constructed from a single majority-like gates. We show that these gates can be used to compute any Boolean function reliably below the noise bound.
\end{abstract}

\begin{IEEEkeywords}
Random Boolean formulas, $\epsilon$-noise, reliable computation.
\end{IEEEkeywords}

\IEEEpeerreviewmaketitle

\section{Introduction}

\IEEEPARstart{O}{ne} of computation models for a Boolean function $\rmf:\{-1,1\}^N\rightarrow\{-1,1\}$  is a Boolean  circuit or formula ~\cite{BooleanBook}. A  \emph{circuit}  is a directed acyclic graph in which nodes of in-degree zero  are either the Boolean constants or variables, nodes of in-degree $k\geq1$ are logical gates, computing Boolean functions  of $k$ arguments, and nodes of out-degree zero  correspond to the circuit outputs. If   a circuit has only a single output and  the output of each gate is used as an input to at most one gate then this circuit  is called a  \emph{formula}. In circuits,  as in any other model of computation,  the computational complexity and effects of noise are  important questions~\cite{Hajnal}.

The circuit complexity of a Boolean function is the minimum number of gates  (circuit \emph{size})  or the minimum  \emph{depth}\footnote{The depth of a circuit  is  the number of gates on the longest path from an input node to the output node.} of a circuit, constructed from a particular set of gates, which computes this function.  However,  to find a circuit representation of a Boolean function  with a bounded  size or depth   is  a difficult problem~\cite{BooleanBook}. One approach to this problem  is to study complexity of \emph{typical} Boolean functions computed by random formulas~\cite{Brodsky}.

The two most studied methods of generating random formulas use random tree generation and a growth process as their core procedures. In the first method, a rooted $k$-ary tree is sampled from the uniform distribution of all rooted $k$-ary trees; the leaves of this random tree are then labelled by reference to the Boolean variables and internal nodes are labelled by the Boolean gates. This method was used to investigate the complexity of typical functions computed by  random AND/OR formulas~\cite{Lefmann,Chauvin,Gardy} and allowed to obtain a close relation between  the probability of a random formula to compute a Boolean function and its size (complexity). However, it seems that this probability distribution is biased towards very low complexity functions~\cite{Chauvin}.

The second method uses the following growth process: Firstly, one defines
an arbitrarily chosen initial probability distribution $\mathcal{P}_{0}$ over the set $\mathcal{F}^0$  of  Boolean functions of $N$ variables. Secondly, and in further steps, the functions chosen from the distributions $\mathcal{P}_{t}$  defined in previous steps are combined by Boolean gates to generate a new set of Boolean functions: $\mathcal{F}^{t+1}=\{\alpha(\rmf_1,\ldots,\rmf_k); \rmf_j\in \mathcal{F}^t \textrm{ for } j=1,2,\ldots,k\}$, where $\alpha:\{-1,1\}^k\rightarrow\{-1,1\}$. This process can be seen as a growth of $k$-ary balanced trees and was first used by Valiant to obtain an upper bound on the size of monotone  formulas computing the majority function~\cite{Valiant}.  Savick\'{y} recently showed for one of these processes, for $\mathcal{P}_{0}$ that is uniform on  some set of Boolean functions $\mathcal{F}^0$ and under very broad conditions on $\alpha$, the probability  $\mathcal{P}_{t}$    tends to the uniform distribution over all Boolean functions of $N$ variables when 
$t\rightarrow\infty$~\cite{Savicky}.  The
convergence rates of the
Savick\'{y}'s process and its variants with different gates and initial conditions were studied in~\cite{Brodsky}.

Another important question in the circuit theory  is a reliable computations of Boolean functions in the presence of noise. One of the first to study the effect of noise in computing systems was von Neumann who attempted to explain the robustness of biologically-inspired computing circuits~\cite{VonNeumann}.  His model represented neural activities by a circuit (or formula) composed of $\epsilon$-noisy Boolean gates. The  $\epsilon$-noisy gate is designed to compute a Boolean function  $\alpha(\sigma)$,  but for each input $\sigma\in\{-1,1\}^k$  there is an error probability $\epsilon$ such that $\alpha(\sigma)\rightarrow-\alpha(\sigma)$. To simplify the analysis, error-probability is taken to be independent for each gate in the circuit. Clearly, a noisy circuit ($\epsilon>0$) cannot perform any given computation in a deterministic manner: for any circuit-input there is a non-vanishing probability that the circuit will produces the wrong output.  The maximum 
of this error probability $\delta$ over all circuit-inputs determines the \emph{reliability} of the circuit.  In his paper, von Neumann showed that reliable computation ($\delta<1/2$) is possible for a sufficiently small $\epsilon$~\cite{VonNeumann} and demonstrated how reliability of a Boolean noisy circuit can be improved by using constructions based only on $\epsilon$-noisy gates.

There had been little development in the analysis of noisy computing systems until the seminal work of Pippenger~\cite{Pippenger:RC} who addressed the problem from an information theory point of view. He showed that if a noisy $k$-ary formula is used to compute a Boolean function $f$ with the error probability $\delta<1/2$, then (i) there is an upper bound for the gate-error $\epsilon(k)$ which is strictly less than $1/2$ and (ii)
there is a lower bound for the formula-depth $\hat{d}(k,\epsilon,\delta)\ge d$, where $d$ is the depth of a noiseless formula computing $f$. In comparison to its noiseless counterpart, a noisy formula that computes reliably has greater depth due to the presence of restitution-gates, implying longer computation times~\cite{Pippenger:RC}.

A number of papers have followed and extended Pippenger's results. For instance, similar results were derived for circuits by Feder~\cite{Feder:RC}, who also improved the bounds obtained by Pippenger for formulas. The exact noise thresholds for $k$-ary Boolean formulae were later determined for odd $k$~\cite{Hajek:MTN,Evans:MTNK} and  for  formulas constructed from $2$-input NAND gates~\cite{Evans:MTN}; the latter was recently suggested as the exact noise threshold for general $2$-input gate formulas~\cite{Unger}.

Results derived for noisy Boolean formulas in~\cite{Hajek:MTN,Evans:MTNK} rely on a specific construction which uses $\epsilon$-noisy  \emph{majority} gates. The noiseless variant of this gate performs the majority-vote function\footnote{We use the definition $\sgn[x]\!=\!1$ for $x>0$, $\sgn[x]\!=\!-1$ for $x<0$ and  $\sgn[0]\!=\!0$ throughout this paper.} $\sgn[\sum_{i=1}^k \sigma_j]$ on the binary inputs $\sigma_j\in\{-1,1\}$ and naturally the number of these inputs $k$ is \emph{odd}. In contrast to previous work, in this paper we concentrate on the possibility of reliable computation in \emph{randomly generated} Boolean formulas. As a first step towards this goal, we study the effects of $\epsilon$-noise on the formulas generated in the Savick\'{y}'s growth process.

\section{Noisy growth processes and main results}
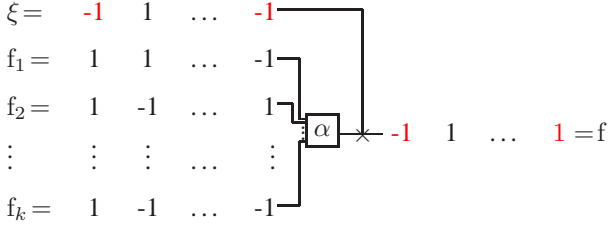
\begin{figure}[!t]

\setlength{\unitlength}{1mm}

\begin{picture}(50,30)

\put(-3.5,12){

\begingroup
\renewcommand*{\arraystretch}{1.5}
\begin{tabular}{ l c c c r }
   $\xi\!=\!$& \textcolor{red}{-1}& 1 & \ldots &\textcolor{red}{-1}\\
  $\rmf_1\!=$& 1& 1 & \ldots &-1\\
    $\rmf_2\!=$ & 1& -1 & \ldots &1 \\
  \vdots & \vdots & \vdots & \ldots &\vdots \\
    $\rmf_k\!=$ & 1 &-1 & \ldots &-1 \\
\end{tabular}
\endgroup
}

\linethickness{0.7pt}

\put(41,8){\framebox(4,4){$\alpha$}}

\put(50,8.5){\begin{tabular}{ l c c r }
   \textcolor{red}{-1}& 1 & \ldots &$\textcolor{red}{1}=\!\rmf$
  \end{tabular}}


\put(37, 19.5){\line(1, 0){3}}
\put(37, 13.5){\line(1, 0){2}}
\put(37, 0){\line(1, 0){3}}

\put(40, 19.5){\line(0, -1){8}}

\put(39, 13.5){\line(0, -1){2.5}}

\put(40, 0){\line(0, 1){8.5}}

\put(40, 11.5){\line(1, 0){1}}

\put(39, 11){\line(1, 0){2}}

\put(40, 8.5){\line(1, 0){1}}

\put(40, 10.2){\small{$.$}}
\put(40, 9.5){\small{$.$}}
\put(40, 8.7){\small{$.$}}

\put(45, 9.5){\line(1, 0){6}}

\put(37, 26){\line(1, 0){11.3}}
\put(48.3, 26){\line(0, -1){16.5}}

\put(47, 8.5){$\times$}
%
\end{picture}
%
\caption{Noisy growth process. i) Boolean functions $\rmf_1,\ldots,\rmf_k$ (represented by binary strings of length $2^N$) are sampled randomly and independently from the distribution $\Prob_t[\,\rmf\,]$. ii) These functions are then used to compute a new Boolean function $\rmf$ via the gate $\alpha$. At each step of this computation noise (represented by the binary string $\xi$) inverts the output of $\alpha$ (this operation is represented by the $\times$ symbol) with probability $\epsilon$. In this figure the first and the last bits of the function $\rmf$ (in red) are inverted by noise. Repeating operations i) and ii) many times gives rise to an ensemble described by the  distribution $\Prob_{t+1}[\,\rmf\,]$.    \label{fig:1} 
 }
\end{figure}

The model we study is given by the following growth process: Starting from any arbitrarily chosen initial probability distribution $\Prob_{0}$ over the set $\rmF^0$ of $N$-variable Boolean functions, one recursively uses functions chosen from the distributions $\Prob_{t}$  defined at a previous step $t$ to determine the new set of Boolean functions at step $t+1$ such that $\rmF^{t+1}=\{\xi(\sigma)\alpha(\rmf_1(\sigma),\ldots,\rmf_k(\sigma)); \rmf_j\in \rmF^t \textrm{ for } j=1,2,\ldots,k\}$, where $\alpha$ is a $k$-ary Boolean gate and $\xi(\sigma)$ is a random Boolean function where for each input $\sigma\in \{-1,1\}^N$ its output is drawn independently and at random with the probability $\Prob(\xi(\sigma)=-1)=\epsilon$. This process can be seen as a noisy version  of  the Savick\'{y}'s growth process (described in the introduction)~\cite{Savicky}. To distinguish the noisy from the noiseless variants of this process we will denote the probability over functions at step $t$ as $\Prob_{t}$ and
the corresponding set of Boolean functions by $\rmF^{t}$  for the former and as $\mathcal{P}_{t}$ and $\mathcal{F}^{t}$ for the latter. 

The growth  process  can be  also seen as a computation, performed by gate $\alpha$, of a new Boolean function\footnote{We index all elements of  $\{-1,1\}^N$ using $i=1,\ldots,2^N$ such that the $i$-th component of $\rmf$, $\rmf^i$, is an output of the function $\rmf:\{-1,1\}^N\rightarrow\{-1,1\}$ for the $i$-th input.}   $\rmf\in\{-1,1\}^{2^N}$  from $k$ Boolean functions $\rmf_1,\ldots,\rmf_k$. These functions $\rmf_j\in\{-1,1\}^{2^N}~\forall j=1,\ldots,k,$  are drawn randomly and independently from the same distribution. However, each computation at the gate $\alpha$  may be corrupted by noise that inverts the result of this computation with  probability $\epsilon$ (see Figure~\ref{fig:1}). Averaging this computation over many  noise realisations leads to the equation

\begin{eqnarray}
\Prob_{t+1}[\,\rmf\,]&=&\sum_{\rmf_1,\ldots,\rmf_k} \bigg\{\prod_{j=1}^k \Prob_t[\,\rmf_j\,]\bigg\}\label{eq:ProbOfF}\\
&&\times\prod_{i=1}^{2^N}\frac{\rme^{\beta \rmf^i\alpha(\rmf_1^i,\ldots,\rmf_k^i)}}{2\cosh(\beta) }~,\nonumber
\end{eqnarray}
where the summation is over all $k$-tuples $(\rmf_1,\ldots,\rmf_k)$ and $f^{i}_{j}$ refers to the output of the Boolean function  $j$ to the $i$-th input. This gives us the probability of a Boolean function $\rmf$ being computed by the noisy formulas of depth $t+1$. Here for convenience we have introduced the inverse ``temperature'' parameter $\beta=1/T$ which is related  to the noise parameter $\epsilon$ via the equality $\epsilon=(1-\tanh\beta)/2=\rme^{-\beta}/2\cosh(\beta)$ ($1-\epsilon=\rme^{\beta}/2\cosh(\beta)$).  The limits $\beta\rightarrow0/\infty$ correspond to completely random/deterministic cases.

Without noise ($\beta\rightarrow\infty$)  Equation~(\ref{eq:ProbOfF}) reduces to
\begin{eqnarray}
\mathcal{P}_{t+1}[\!\;\rmf\!\;]&=&\sum_{\rmf_1,\ldots,\rmf_k}\bigg\{\prod_{j=1}^k \mathcal{P}_t[\;\rmf_j\;]\bigg\}\label{eq:ProbOfFnoNoise}\\
&&\times\prod_{i=1}^{2^N} \delta\left[\rmf^i;\alpha(\rmf_1^i,\ldots,\rmf_k^i)\right],\nonumber
\end{eqnarray}
where we use $\delta[x;y]$ to denote Kronecker delta. Equation~(\ref{eq:ProbOfFnoNoise}) was studied in the original Savick\'{y}'s work~\cite{Savicky} and subsequent studies~\cite{Brodsky} where the stationary distribution  $\mathcal{P}_\infty[\!\;\rmf\!\;]=\lim_{t\rightarrow\infty}\mathcal{P}_t[\!\;\rmf\!\;]$ of the noiseless process (\ref{eq:ProbOfFnoNoise}) was studied with the initial conditions $\mathcal{P}_{0}[\!\;\rmf\!\;]\;\;\;=\frac{1}{\vert\mathcal{F}^0\vert}\sum_{\rmg\in\mathcal{F}^0}\prod_{i=1}^{2^N}\delta[\rmf^i;\rmg^i] $ for  different initial sets $\mathcal{F}^0$  of simple Boolean functions (constants, identities, etc.) and different gates $\alpha$. Depending on these parameters the stationary distribution is  either concentrated on a single function, i.e. $\mathcal{P}_\infty[\!\;\rmf\!\;]=\prod_{i=1}^{2^N}\delta\left[\;\rmf^i;\rmg^i\right]$ or on some set of functions $\mathcal{F}$, i.e. $\mathcal{P}_\infty[\!\;\rmf\!\;]=  \frac{1}{\vert\mathcal{F}\vert}\sum_{\rmg\in\mathcal{F}}\prod_{i=1}^{
2^N}\delta[\rmf^i;\rmg^i]$. There are also
cases when for $t\rightarrow\infty$ the distributions $\mathcal{P}_t[\!\;\rmf\!\;]$ and $\mathcal{P}_{t+1}[\!\;\rmf\!\;]$ are distinct.

Our main contribution to these studies is the following result for the recursion relation  (\ref{eq:ProbOfF}).
\begin{theorem}\label{theorem:1}
For \emph{any} initial distribution  $\Prob_0[\,\rmf\,]$ and  \emph{balanced} gate\footnote{The gate is balanced when over all input vectors it has an equal number of $+1$'s and $-1$'s in its output.} $\alpha$  the stationary distribution  $\Prob_{\infty}[\rmf]=\frac{1}{2^{2^N}}$ is  the unique and stable solution of the recursion relation (\ref{eq:ProbOfF}) when $\epsilon>\epsilon(k)=\frac{1-b(k)}{2}$, where $b(k)\!\equiv \!\left\{2^{k\!-\!1}/k\binom{k\!-\!1}{(k\!-\!1)/2};\; 2^{k\!-\!2}/(k\!-\!1)\binom{k\!-\!2}{(k\!-\!2)/2}\right\}$, with $k\geq3$, for $k$ odd and even respectively.
\end{theorem}
\begin{IEEEproof}
In order to show this, we employ three lemmas. First we use a well known fact that:
 \begin{lemma}\label{lemma:moments1}
 The distribution $\Prob_{t+1}[\,\rmf\,]$ can be represented via its moments $\rmm^S(t+1)=\sum_{\hat\rmf}\Prob_{t+1}[\,\hat\rmf\,]\prod_{i\in S}\hat\rmf^i$, where $S$ is a subset of the set $[2^N]=\{1,\ldots,2^N\}$,   and $\Prob_{t+1}[\,\rmf\,]$ is given by
\begin{eqnarray}
\Prob_{t+1}[\,\rmf\,]=\frac{1}{2^{2^N}}\left(1+\sum_{S\subseteq[2^N] \setminus\emptyset}\rmm^S(t+1)\prod_{i\in S}\rmf^i\right),
\end{eqnarray}
\end{lemma}
See Appendix~\ref{section:m} for the proof.

We then employ the following lemma:
 \begin{lemma}\label{lemma:moments2} The $n$-th moment of the distribution $\Prob_{t+1}[\,\rmf\,]$ is  governed by the equation
\begin{eqnarray}
&&\rmm^\setI(t+1)\label{eq:p-m}\\
&&=\tanh^n(\beta)
%
\sum_{\rmf_1^{i_1},\ldots,\rmf_1^{i_n}} \cdots \sum_{\rmf_k^{i_1}   ,\ldots,\rmf_k^{i_n}}\nonumber\\
&&\times\prod_{j=1}^k \frac{1}{2^{n}}\left[1+\sum_{S\subseteq\setI\setminus\emptyset}\rmm^S(t)\prod_{i\in S}\rmf^i_j\right]\nonumber\\
&&\times\prod_{i\in \setI}\alpha(\rmf_1^{i},\ldots,\rmf_k^{i}),\nonumber
\end{eqnarray}
where $\setI=\{i_1,\ldots,i_n\}$.
\end{lemma}
For the proof see Appendix~\ref{section:m2}. From this lemma follows that  the $n$-th moment at $t+1$ is a function of only the $n$-th and lower order moments at $t$.

Let us now introduce the following lemma:
\begin{lemma}\label{lemma:moments3}
Suppose $\alpha$ is a balanced gate and assume that all moments but the $n$-th  vanish,  then the point $\rmm=0$ is a stable and unique solution of (\ref{eq:p-m}) when $\tanh^n(\beta)<b(k)$.
\end{lemma}
For the proof see Appendix~\ref{section:moments}.

Using this lemma for $n=1$, the first moments of the distribution $\Prob_{t+1}[\,\rmf\,]$ vanish as $t\rightarrow\infty$. But then, by applying the same lemma to the orders $n\geq2$ moments, we conclude that all moments are vanishing  as $t\rightarrow\infty$. Hence $\Prob_{\infty}[\rmf]=\frac{1}{2^{2^N}}$ represents the unique and stable solution of the recursion Equation~(\ref{eq:ProbOfF}).
\end{IEEEproof}

\begin{figure}
\setlength{\unitlength}{1.0mm}
\begin{picture}(50,50)
\put(0,0){\epsfysize=50\unitlength\epsfbox{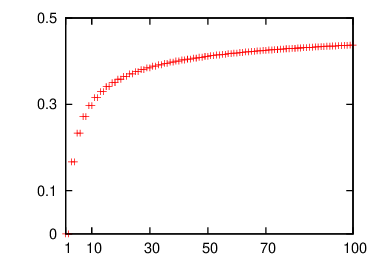}}
\put(1,35){$\epsilon(k)$}
\put(39,-3){$k$}
\end{picture}
\caption{Upper bound for reliable computation by noisy $k$-ary random formulas. \label{fig:2}}
\end{figure}

In addition to its direct interpretation that above $\epsilon(k)$ (see Figure~\ref{fig:2}) the noisy process~(\ref{eq:ProbOfF}) is \emph{ergodic} and has only one stationary solution, the result of Theorem~\ref{theorem:1}  also has consequences for computation  in noisy random formulas.  A feature of noisy formulas, which is  essential for reliable computation, is their greater depth due to the presence of correcting $\epsilon$-noisy gates~\cite{Pippenger:RC}. This correction operation can be seen as a procedure which reduces the entropy, but in our case of very deep ($t\rightarrow\infty$) random formulas the entropy is at its maximum when $\epsilon>\epsilon(k)$. Thus any computation, even as simple as computing the identity function, can not be performed reliably in this regime.

For odd $k$ our result for the bound $\epsilon(k)$ is exactly equal to the  \emph{exact} threshold\footnote{Notice that as the index $i$ runs over the same (all) input choices for all $k$ Boolean functions $\rmf_1,\ldots,\rmf_k$ in Equation~(\ref{eq:ProbOfF}), hence the gate entries used to generate new functions are not statistically independent and our result cannot be directly mapped onto the framework of~\cite{Evans:MTNK}.} for reliable computation by general $k$-ary formulas~\cite{Hajek:MTN,Evans:MTNK}. It is not clear however if this threshold is also exact, i.e. any Boolean function can be computed for $\epsilon\in(0,\epsilon(k))$ with the error $\delta<1/2$,
 for randomly generated formulas. For even $k>2$ this threshold is not known, but our result suggests that for balanced gates $\alpha$ it can not exceed the bound $\epsilon(k)$ of Theorem~\ref{theorem:1}. Furthermore as  $k\rightarrow\infty$ the $\epsilon(k)$ approaches $1/2$ as $1/2-\epsilon(k)=O(1/\sqrt{k})$, this follows from the Stirling'
s approximation of $b(k)$, which is in agreement with the bound computed in~\cite{Evans:SPNC} for general formulas.

\section{Computation of the lower bound values\label{section:ub}}
In this section we compute the values of lower bounds appearing in  Theorem \ref{theorem:1}.
In order to do this we  choose a balanced gate $\chi(\sigma_1,\ldots,\sigma_k)$ from the set of gates $\sgn\left[\sum_{j=1}^k \sigma_j\right]+\1\left[\sum_{j=1}^k \sigma_j=0\right]\gamma(\sigma_1,\ldots,\sigma_k)$, where $\gamma(\sigma_1,\ldots,\sigma_k)\in\{\!-\!1,1\}$ is such that $\sum_{\sigma_1,\ldots,\sigma_k}\1\left[\sum_{j=1}^k \sigma_j=0\right]\gamma(\sigma_1,\ldots,\sigma_k)\!=\!0$. The input variables $\sigma_j \in\{-1,1\}$ represent arbitrary binary inputs.  This gate can be seen as a generalisation of the majority gate (for even $k$) performing majority-vote function when more than half of its inputs are $+1$ (or $-1$) and providing a
balanced output otherwise. Also, this construction satisfies conditions of Savick\'{y}'s growth process~\cite{Savicky}.

Let us now consider  the first moments $\rmm_i(t)=\sum_{\rmf}\Prob_{t}[\,\rmf\,]\,\rmf^i$ of the distribution (\ref{eq:ProbOfF})  where as a specific choice we employ the Boolean gate $\chi$ such that $\alpha\equiv\chi$. These are governed by the equations (derived in  Appendix~\ref{section:F})
\begin{eqnarray}
\rmm(t+1)&=& F_{\chi}^1(\rmm(t))\label{eq:m}\\
F_{\chi}^1(\rmm)&=&\tanh(\beta)\sum_{\ell=0}^k \binom{k}{\ell}\sgn[2\ell-k]\nonumber \\
&&\times\left[\frac{1+\rmm}{2}\right]^{\ell}\left[\frac{1-\rmm}{2}\right]^{k-\ell}\nonumber,
\end{eqnarray}
where for $\rmm(t)=\pm1$  we use $0^0=1$.
\begin{lemma}\label{lemma:1}
For $k\geq3$ the function $F_\chi^1(\rmm)$ has the following properties: i) if $\tanh\beta\leq b(k)$ then $\rmm>F_\chi^1(\rmm)$ for $\rmm\in(0,1]$ and $F_\chi^1(\rmm)>\rmm$ for $\rmm\in[-1,0)$; ii) if $\tanh\beta> b(k)$ then $ \exists\; \rmm^*\neq 0$ such that $\rmm^*=F_\chi^1(\rmm^*)$,
where $b(k)$ is defined in Theorem~\ref{theorem:1}.
\end{lemma}
\begin{IEEEproof}
This lemma follows from  the equalities $F_\chi^1(\pm1)=\pm\tanh\beta, F_\chi^1(0)=0$ (this can be shown by direct substitution) and the fact that $F_\chi^1(\rmm)$ is a strictly increasing function, which is also convex and concave on the intervals $(-1,0)$ and $(0,1)$, respectively (to show this we study properties of $F_\chi^1(\rmm)$ in Appendix~\ref{section:F}). Then i) is true because $\frac{\rmd F_\chi^1}{\rmd\rmm}\vert_{\rmm=0}<1$ when $\tanh\beta< b(k)$ and  ii) is true because  of $\frac{\rmd F_\chi^1}{\rmd\rmm}\vert_{\rmm=0}\geq1$ when $\tanh\beta\geq b(k)$.
\end{IEEEproof}
The results of Lemma~\ref{lemma:1} can be  used to show that reliable computation in randomly generated formulas is possible.
\begin{corollary}\label{corollary:1}
Suppose that $\alpha$ in the  recursive Equation~(\ref{eq:ProbOfF}) is  a generalized majority-vote gate and assume that
the arbitrary initial distribution $\Prob_{0}[\,\rmf\,]$ for this equation  is such that the stationary distribution $\mathcal{P}_{\infty}[\,\rmf\,]$ of the (noiseless) recursion (\ref{eq:ProbOfFnoNoise}), with $\mathcal{P}_{0}[\,\rmf\,]=\Prob_{0}[\,\rmf\,]$ , has only \emph{one} Boolean function in its support. Then \emph{on average} this  Boolean function can be computed with any desired accuracy when  $\epsilon<\epsilon(k)$.
\end{corollary}
\begin{IEEEproof}
The hypothesis assumes  that \emph{without} noise \emph{all} random formulas compute  \emph{the same} Boolean function $\hat\rmf$. Then in the presence of noise, due to  $\hat\rmf^i=\sgn[\rmm_i(\infty)]$,  the \emph{average} formula errors in its output  occur with probability $\mathrm{Prob}(\rmf^i\neq\hat\rmf^i) =\sum_\rmf\Prob_{\infty}[\,\rmf\,]\,\delta[\rmf^i\hat\rmf^i;-1] =(1-\vert\rmm_i(\infty)\vert)/2$, where $\rmm_i(\infty)$ is the stationary solution of Equation~(\ref{eq:m}) corresponding to the $i$-th input. By Lemma~\ref{lemma:1} this error is bounded  below $1/2$ when  $\epsilon<\epsilon(k)$ ($\tanh\beta>b(k)$). Furthermore, it can be reduced by decreasing  $\epsilon$ (the magnitude of $F_\chi^1(\rmm)$ is controlled by  $\tanh\beta=1-2\epsilon$) or by increasing $k$ ($F_\chi^1(\rmm)$ is a monotone increasing  function of $k$ when  $\epsilon<\epsilon(k)$). Thus in this regime  a Boolean function $\hat\rmf$ can
be computed with any desired accuracy.
\end{IEEEproof}

\section{Conclusion}
The paper extends previous work~\cite{Hajek:MTN,Evans:MTNK} on the reliability of computation in Boolean formulas and generation of random Boolean functions~\cite{Savicky,Brodsky}, by investigating the properties of formulas constructed by a random growth process whereby computing elements, primarily $k$-ary balanced gates, are subject to $\epsilon$-noise.

We show that the noisy growth process is ergodic above the noise bound $\epsilon(k)$ and hence the formulas generated by it are unreliable. We also show that  formulas constructed from majority-like gates, which saturate this bound, can be used for computing any Boolean function when $\epsilon<\epsilon(k)$. Our earlier work, which uses methods of non-equilibrium statistical physics, suggests that the same noise bound  also applies to the noisy feed-forward~\cite{noisyPRE} and recurrent Boolean networks~\cite{PhilMag}.

The current analysis  is restricted to reliable computation in a growth process that uses only \emph{balanced} gates\footnote{The results of this paper can be easily extended to the \emph{distributions} over  balanced gates~\cite{PhilMag}.} and produces (without noise) only \emph{one} Boolean function;  but we envisage that it can be extended to study more general scenarios of non-balanced gates and a  richer distributions of Boolean functions~\cite{Brodsky}.

\appendices

\section{Moment representation of $\Prob_{t}[\;\rmf\;]$- Proof of Lemma~\ref{lemma:moments1}\label{section:m}}
\begin{IEEEproof}
The probability distribution $\Prob_{t}[\,\rmf\,]$ can be represented via its moments.  In order to find this representation we can use the identity\footnote{ This  follows from $\delta[x;y]=\frac{1}{2}(1+xy)$ for $x,y\in\{-1,1\}$.  } $\sum_{\hat\rmf}\delta[\hat\rmf;\rmf]=1=\sum_{\hat\rmf}\prod_{i=1}^{2^N}\frac{1}{2}(1+\hat\rmf^i  \rmf^i )$ to write $\Prob_t[\,\rmf\,]=\sum_{\hat\rmf}\delta[\hat\rmf;\rmf]\Prob_t[\,\hat\rmf\,]$. Then  we obtain
\begin{eqnarray}
\Prob_t[\rmf]&=&\sum_{\hat\rmf}\Prob_t[\hat\rmf]\prod_{i=1}^{2^N}\frac{1}{2}\left(1+\hat\rmf^i\rmf^i\right)\label{def:ProbOfFviAm}\\
&=&\frac{1}{2^{2^N}}\left(1+\sum_{\hat\rmf}\Prob_t[\hat\rmf]\sum_{S\subseteq[2^N]\setminus\emptyset}\prod_{i\in S}\hat\rmf^i\rmf^i\right)\nonumber\\
&=&\frac{1}{2^{2^N}}\left(1+\sum_{S\subseteq[2^N]\setminus\emptyset}\rmm^S(t)\prod_{i\in S}\rmf^i\right),\nonumber
\end{eqnarray}
where $\rmm^S(t)=\sum_{\hat\rmf}\Prob_t[\,\hat\rmf\,]\prod_{i\in S}\hat\rmf^i$ ($\rmm^S(t)\in[-1,1]$) are the moments of $\Prob_t[\,\rmf\,]$.
\end{IEEEproof}

\section{Moments of $\Prob_{t}[\;\rmf\;]$ - Proof of Lemma~\ref{lemma:moments2} }\label{section:m2}
\begin{IEEEproof}
Let us now derive an explicit expression for the $n$-th moment of the distribution (\ref{eq:ProbOfF}). This can be obtained by multiplying both sides of  Equation~(\ref{eq:ProbOfF}) by the monomial $\prod_{i\in \setI}\rmf^i$, where $\setI=\{i_1,\ldots,i_n\}$, and taking the sums over $\rmf$ as follows
\begin{eqnarray}
\rmm^\setI(t+1)&=&\sum_\rmf\Prob_{t+1}[\!\;\rmf\!\;]\prod_{i\in \setI}\rmf^i\label{def:m}\\
&&=\sum_\rmf\sum_{ \rmf_1,\ldots,\rmf_k}\prod_{j=1}^k \Prob_t[\;\rmf_j\;]\nonumber\\
&&\times\prod_{i=1}^{2^N}\frac{\rme^{\beta \rmf^i\;\!\alpha(\rmf_1^i,\ldots,\rmf_k^i)}}{2\cosh(\beta) }\prod_{\ell\in \setI}\rmf^\ell\nonumber\\
&&=\sum_{\rmf_1^{1},\ldots,\rmf_1^{2^N}} \cdots \sum_{\rmf_k^{1}   ,\ldots,\rmf_k^{2^N}}\prod_{j=1}^k \Prob_t\left[\rmf_j^1,\ldots,\rmf_j^{2^N}\right ]\nonumber\\
&&\times\prod_{i\in \setI}\sum_{\rmf^i}\frac{\rme^{\beta \rmf^i\;\!\alpha(\rmf_1^i,\ldots,\rmf_k^i)}}{2\cosh(\beta) }\rmf^i\nonumber\\
&&=\tanh^n(\beta)
\sum_{\rmf_1^{i_1},\ldots,\rmf_1^{i_n}} \cdots \sum_{\rmf_k^{i_1}   ,\ldots,\rmf_k^{i_n}}\nonumber\\
%
%
&&\times\prod_{j=1}^k \Prob_t[\;\rmf_j^{i_1},\ldots,\rmf_j^{i_n}\;]\nonumber\\
&&\times\prod_{i\in \setI}\alpha(\rmf_1^{i},\ldots,\rmf_k^{i}),\nonumber
\end{eqnarray}
where in the above $\Prob_t[\,\rmf_j^{i_1},\ldots,\rmf_j^{i_n}\,] $ is a marginal of $\Prob_{t}[\,\rmf_j\, ]$  and we have used the identity $\sum_x  \frac{        \rme^{\beta x y}     }{   2\cosh(\beta)    }x=y\tanh(\beta)$ which is valid for $x,y\in\{-1,1\}$. Finally, using the moment representation of $\Prob_t[\,\rmf_j^{i_1},\ldots,\rmf_j^{i_n}\,]=\frac{1}{2^{n}}\left[1+\sum_{S\subseteq\setI\setminus\emptyset}\rmm^S(t)\prod_{i\in S}\rmf^i_j\right]$, we obtain
\begin{eqnarray}
\rmm^\setI(t+1)&=&\tanh^n(\beta)
\sum_{\rmf_1^{i_1},\ldots,\rmf_1^{i_n}} \cdots \sum_{\rmf_k^{i_1}   ,\ldots,\rmf_k^{i_n}}\label{eq:n-th-moment}\\
&&\times\prod_{j=1}^k \frac{1}{2^{n}}\left[1+\sum_{S\subseteq\setI\setminus\emptyset}\rmm^S(t)\prod_{i\in S}\rmf^i_j\right]\nonumber\\
&&\times\prod_{i\in \setI}\alpha(\rmf_1^{i},\ldots,\rmf_k^{i}).\nonumber
\end{eqnarray}
\end{IEEEproof}
%

\section{Analysis of moments of $\Prob_{t}[\;\rmf\;]$- Proof of Lemma~\ref{lemma:moments3}\label{section:moments}                 }
\begin{IEEEproof}
Let us consider  Equation~(\ref{eq:p-m}) for an $n$-th moment $\rmm$. Assuming that all lower order moments  vanish allows us to write this equation in a very simple form
\begin{eqnarray}
\rmm(t+1)&=& F_\alpha^n(\rmm(t))\label{eq:p-m-reduced}\\
&=&\tanh^n(\beta)
%
%
\sum_{\sigma_1^1,\ldots ,\sigma_1^n}\cdots \sum_{\sigma_k^1,\ldots ,\sigma_k^n            } \nonumber\\
&&\times \prod_{j=1}^k\frac{1}{2^{n}}\left[1+\rmm(t)\prod_{i=1}^n\sigma^i_j\right]\nonumber\\
&&\times\prod_{i=1}^n\alpha(\sigma_1^{i},\ldots,\sigma_k^{i}),\nonumber
\end{eqnarray}
where $F_\alpha^n(\rmm(t))$ represents the $n$-th moment of the distribution obtained for a growth process at step $t+1$ and any balanced gate $\alpha$. For a balanced gate $\alpha$  the point $\rmm=0$ is a stable and unique solution of (\ref{eq:p-m-reduced}) when $\tanh^n(\beta)<b(k)$.

In order to prove this we first show that
\begin{eqnarray}
&&\tanh(\beta)^{n-1}\rmF_\chi^1(m)\\
&&=\tanh^n(\beta)
\sum_{\sigma_1^1,\ldots ,\sigma_1^n}\cdots \sum_{\sigma_k^1,\ldots ,\sigma_k^n            } \nonumber\\
&&\times   \left\{    \prod_{j=1}^k \frac{1}{2^{n}}\left[1+\rmm\prod_{i=1}^n\sigma^i_j\right]    \right\}           \sgn\left[\sum_{j=1}^k\prod_{i=1}^n\sigma^i_j\right],              \nonumber
\end{eqnarray}
where $F_\chi^1(m)$ is defined in (\ref{eq:m}). This can be shown by a direct calculation as follows
\begin{eqnarray}
&&\tanh^n(\beta) \sum_{\sigma_1^1,\ldots ,\sigma_1^n}\cdots \sum_{\sigma_k^1,\ldots ,\sigma_k^n            } \\
&&~~~~\times   \left\{        \prod_{j=1}^k \frac{1}{2^{n}}\left[1+\rmm\prod_{i=1}^n\sigma^i_j\right]  \right\}   \sgn\left[\sum_{j=1}^k\prod_{i=1}^n\sigma^i_j\right]       \nonumber           \\
&&=\tanh^n(\beta) \sum_{\sigma_1}\cdots\sum_{\sigma_k}\nonumber\\
&&~~~~\times\prod_{j=1}^k\left\{\sum_{\sigma_j^1,\ldots ,\sigma_j^n}\frac{1}{2}\left[1+\sigma_j\prod_{i=1}^n\sigma^i_j \right]\right\}\nonumber \\
&&~~~~\times   \left\{        \prod_{j=1}^k \frac{1}{2^{n}}\left[1+\rmm\,\sigma_j\right]  \right\}                  \sgn\left[\sum_{j=1}^k\sigma_j\right]       \nonumber           \\
&&= \tanh^n(\beta)
\sum_{\sigma_1}\cdots \sum_{\sigma_k        }\nonumber\\
&&~~~~\times     \left\{   \prod_{j=1}^k \left[\frac{1+\rmm\,\sigma_j}{2}\right]  \right\}                  \sgn\left[\sum_{j=1}^k\sigma_j\right]\nonumber\\
&&=\tanh(\beta)^{n-1}\rmF_\chi^1(m).\nonumber
\end{eqnarray}
In the above the second equality was obtained by using the identity 
\begin{eqnarray}
&&\prod_{j=1}^k\left\{\sum_{\sigma_j}\delta\left[\sigma_j;\prod_{i=1}^n\sigma^i_j \right]\right\}\\
&&=\prod_{j=1}^k\left\{\sum_{\sigma_j}\frac{1}{2}\left[1+\sigma_j\prod_{i=1}^n\sigma^i_j \right]\right\}=1\nonumber
\end{eqnarray}
and the last equality was obtained by computing the sums and comparing with the equation~(\ref{eq:m}).

Next, for a balanced gate $\alpha$  we compute the difference $\Delta(\rmm)=\tanh^{n-1}(\beta)\rmF_\chi^1(\rmm)-\rmF_\alpha^n(\rmm)$  as follows:
\begin{eqnarray}
&&\frac{\Delta(\rmm)}{4\tanh^n(\beta)}=\sum_{\sigma_1^1,\ldots ,\sigma_1^n}\cdots \sum_{\sigma_k^1,\ldots ,\sigma_k^n            }\label{eq:deltaM1}\\
&&\times   \prod_{j=1}^k \frac{1}{2^n}\left[1+\rmm\prod_{i=1}^n\sigma^i_j\right] \nonumber\\
&&\times \frac{1}{4}\left\{\sgn\left[\sum_{j=1}^k\prod_{i=1}^n\sigma^i_j\right] -\prod_{i=1}^n\alpha(\sigma_1^{i},\ldots,\sigma_k^{i})\right\}\nonumber\\
&&=\sum_{\sigma_1^1,\ldots ,\sigma_1^n}\cdots \sum_{\sigma_k^1,\ldots ,\sigma_k^n            }\prod_{j=1}^k \frac{1}{2^n}\left[1+\rmm\prod_{i=1}^n\sigma^i_j\right] \nonumber\\
&&\times\frac{1}{4}\Bigg\{ \1\left[\sum_{j=1}^k\prod_{i=1}^n\sigma^i_j\!>\!0\right]-\1\left[\sum_{j=1}^k\prod_{i=1}^n\sigma^i_j\!<\!0\right]\nonumber\\
&&-\Bigg( \1\left[\sum_{j=1}^k\prod_{i=1}^n\sigma^i_j\!>\!0\right]   +   \1\left[\sum_{j=1}^k\prod_{i=1}^n\sigma^i_j\!<\!0\right]\nonumber\\
&&   +      \1\left[\sum_{j=1}^k\prod_{i=1}^n\sigma^i_j\!=\!0\right]     \Bigg)\prod_{i=1}^n\alpha(\sigma_1^{i},\ldots,\sigma_k^{i})\Bigg\}\nonumber\\
&&=\sum_{\sigma_1^1,\ldots ,\sigma_1^n}\cdots \sum_{\sigma_k^1,\ldots ,\sigma_k^n            }\prod_{j=1}^k \frac{1}{2^n}\left[1+\rmm\prod_{i=1}^n\sigma^i_j\right] \nonumber\\
&&\times\frac{1}{2}\Bigg\{
 \1\left[\sum_{j=1}^k\prod_{i=1}^n\sigma^i_j\!>\!0\right]\1\left[\prod_{i=1}^n\alpha(\sigma_1^{i},\ldots,\sigma_k^{i})\!=\!-\!1\right]\nonumber\\
&&-\1\left[\sum_{j=1}^k\prod_{i=1}^n\sigma^i_j\!<\!0\right]\1\left[\prod_{i=1}^n\alpha(\sigma_1^{i},\ldots,\sigma_k^{i})\!=\!+\!1\right]\nonumber\\
&&-\frac{1}{2}     \1\left[\sum_{j=1}^k\prod_{i=1}^n\sigma^i_j\!=\!0\right]      \prod_{i=1}^n\alpha(\sigma_1^{i},\ldots,\sigma_k^{i})\Bigg\}.\nonumber
\end{eqnarray}
In the above we can use the identity
\begin{eqnarray}
&&\prod_{j=1}^k \frac{1}{2^n}\left[1+\rmm\prod_{i=1}^n\sigma^i_j\right]=\\
&&\left[\frac{1\!+\!\rmm}{2^n}\right]^{\frac{k+\sum_{j=1}^k\prod_{i=1}^n\sigma^i_j}{2}}\left[\frac{1\!-\!\rmm}{2^n}\right]^{\frac{k-\sum_{j=1}^k\prod_{i=1}^n\sigma^i_j}{2}}\nonumber
\end{eqnarray}
to obtain
\begin{eqnarray}
&&\frac{\Delta(\rmm)}{4\tanh^n(\beta)}\label{eq:deltaM2}\\
&&=\frac{1}{2}\left(\left[\frac{1\!+\!\rmm}{2^n}\right]\left[\frac{1\!-\!\rmm}{2^n}\right]\right)^{\frac{k}{2}}\nonumber\\
&&\times\Bigg\{\sum_{\sigma_1^1,\ldots ,\sigma_1^n}\cdots \sum_{\sigma_k^1,\ldots ,\sigma_k^n            }\left[\frac{1\!+\!\rmm}{1\!-\!\rmm}\right]^{\frac{\vert\sum_{j=1}^k\prod_{i=1}^n\sigma^i_j\vert}{2}}\nonumber\\
&&\times \1\left[\sum_{j=1}^k\prod_{i=1}^n\sigma^i_j\!>\!0\right]\1\left[\prod_{i=1}^n\alpha(\sigma_1^{i},\ldots,\sigma_k^{i})\!=\!-\!1\right]\nonumber\\
&&-\sum_{\sigma_1^1,\ldots ,\sigma_1^n}\cdots \sum_{\sigma_k^1,\ldots ,\sigma_k^n            }\left[\frac{1\!-\!\rmm}{1\!+\!\rmm}\right]^{\frac{\vert\sum_{j=1}^k\prod_{i=1}^n\sigma^i_j\vert}{2}}\nonumber\\
&&\times\1\left[\sum_{j=1}^k\prod_{i=1}^n\sigma^i_j\!<\!0\right]\1\left[\prod_{i=1}^n\alpha(\sigma_1^{i},\ldots,\sigma_k^{i})\!=\!+\!1\right]\nonumber\\
&&-\frac{1}{2}\sum_{\sigma_1^1,\ldots ,\sigma_1^n}\cdots \sum_{\sigma_k^1,\ldots ,\sigma_k^n            }          \1\left[\sum_{j=1}^k\prod_{i=1}^n\sigma^i_j\!=\!0\right]      \nonumber\\
&&\times \prod_{i=1}^n\alpha(\sigma_1^{i},\ldots,\sigma_k^{i})\Bigg\}.\nonumber
\end{eqnarray}
Now because $\alpha$ is a balanced gate we have the following identity
\begin{eqnarray}
&& \sum_{\sigma_1^1,\ldots ,\sigma_1^n}\cdots \sum_{\sigma_k^1,\ldots ,\sigma_k^n            }\prod_{i=1}^n\alpha(\sigma_1^{i},\ldots,\sigma_k^{i})\label{eq:zero}\\
&&=\sum_{\sigma_1^1,\ldots ,\sigma_1^n}\cdots \sum_{\sigma_k^1,\ldots ,\sigma_k^n            }\Bigg( \1\left[\sum_{j=1}^k\prod_{i=1}^n\sigma^i_j\!>\!0\right]\nonumber\\
&&+\1\left[\sum_{j=1}^k\prod_{i=1}^n\sigma^i_j\!<\!0\right]+               \1\left[\sum_{j=1}^k\prod_{i=1}^n\sigma^i_j\!=\!0\right]              \Bigg)\nonumber\\
&&\times\Bigg(\1\left[ \prod_{i=1}^n\alpha(\sigma_1^{i},\ldots,\sigma_k^{i})\!=\!+\!1\right] \nonumber\\
&&-\1\left[ \prod_{i=1}^n\alpha(\sigma_1^{i},\ldots,\sigma_k^{i})\!=\!-\!1\right]\Bigg )\nonumber\\
&&=\sum_{\sigma_1^1,\ldots ,\sigma_1^n}\cdots \sum_{\sigma_k^1,\ldots ,\sigma_k^n            }\Bigg(\nonumber\\
&&\frac{1}{2}     \1\left[\sum_{j=1}^k\prod_{i=1}^n\sigma^i_j\!=\!0\right]          \prod_{i=1}^n\alpha(\sigma_1^{i},\ldots,\sigma_k^{i})\nonumber\\
&&+\1\left[\sum_{j=1}^k\prod_{i=1}^n\sigma^i_j\!<\!0\right]\1\left[\prod_{i=1}^n\alpha(\sigma_1^{i},\ldots,\sigma_k^{i})\!=\!+\!1\right]\nonumber\\
&&- \1\left[\sum_{j=1}^k\prod_{i=1}^n\sigma^i_j\!>\!0\right]\1\left[\prod_{i=1}^n\alpha(\sigma_1^{i},\ldots,\sigma_k^{i})\!=\!-\!1\right]\Bigg)\nonumber\\
&&=0.\nonumber
\end{eqnarray}

 Using the above identity inside the curly brackets in Equation~(\ref{eq:deltaM2}) leads  to the final result
\begin{eqnarray}
&&\Delta(\rmm)\\
&&=2\tanh^n(\beta)\left(\left[\frac{1\!+\!\rmm}{2^n}\right]\left[\frac{1\!-\!\rmm}{2^n}\right]\right)^{\frac{k}{2}}\Bigg\{\nonumber\\
&&\times\sum_{\sigma_1^1,\ldots ,\sigma_1^n}\cdots \sum_{\sigma_k^1,\ldots ,\sigma_k^n            }
\left(\left[\frac{1\!+\!\rmm}{1\!-\!\rmm}\right]^{\frac{\vert\sum_{j=1}^k\prod_{i=1}^n\sigma^i_j\vert}{2}}-1\right)\nonumber\\
&&~~~~\times\1\left[\sum_{j=1}^k\prod_{i=1}^n\sigma^i_j\!>\!0\right]\1\left[\prod_{i=1}^n\alpha(\sigma_1^{i},\ldots,\sigma_k^{i})\!=\!-\!1\right]\nonumber\\
&&+\sum_{\sigma_1^1,\ldots ,\sigma_1^n}\cdots \sum_{\sigma_k^1,\ldots ,\sigma_k^n            }
\left(1-\left[\frac{1\!-\!\rmm}{1\!+\!\rmm}\right]^{\frac{\vert\sum_{j=1}^k\prod_{i=1}^n\sigma^i_j\vert}{2}}\right)\nonumber\\
&&\times\1\left[\sum_{j=1}^k\prod_{i=1}^n\sigma^i_j\!<\!0\right]\1\left[\prod_{i=1}^n\alpha(\sigma_1^{i},\ldots,\sigma_k^{i})\!=\!+\!1\right]\Bigg\}.\nonumber
\end{eqnarray}
 The result of this computation is that $\Delta(\rmm)\geq0$ and $\Delta(\rmm)\leq0$ on the intervals $\rmm\!\in\![0,1)$ and $\rmm\!\in\!(-1,0]$, respectively, from which the bounds $\tanh^{n-1}(\beta)\rmF_\chi^1(\rmm)\geq \rmF_\alpha^n(\rmm)$ and $\tanh^{n-1}(\beta)\rmF_\chi^1(\rmm)\leq \rmF_\alpha^n(\rmm)$ on the same intervals follow. The behavior of $\tanh^{n-1}(\beta)\rmF_\chi^1(\rmm)$ with respect to the inverse temperature $\beta$ is the same as of $\rmF_\chi^1(\rmm)$, which we described in Lemma \ref{lemma:1}, but with the $\tanh(\beta)$ being replaced by the $\tanh^n(\beta)$. This implies that for $\tanh^{n}(\beta)<b(k)$ we have that   $m>\tanh^{n-1}(\beta)\rmF_\chi^1(\rmm)\geq \rmF_\alpha^n(\rmm)$ on the interval $\rmm\!\in\![0,1)$ and $m<\tanh^{n-1}(\beta)\rmF_\chi^1(\rmm)\leq \rmF_\alpha^n(\rmm)$ on the interval  $\rmm\!\in\!(-1,0]$. Now $\rmF_\alpha^n(0)=0$ and 
hence $\rmm=0$ is a stable and unique solution of the recursion (\ref{eq:p-m-reduced}). 
\end{IEEEproof}

\section{Derivation of $F_\chi^1$ and analysis of its properties\label{section:F}}
Here we compute the function $F_\chi^1(\rmm)$  and study its properties. Let us first compute the sum
\begin{eqnarray}
&&F_\chi^1(\rmm)\nonumber\\
&&~~ = \sum_{  \sigma_1,\ldots,\sigma_k }      \prod_{j=1}^k \left[\frac{1+\sigma_j\;\rmm}{2}\right] \Bigg\{\sgn\left[\sum_{j=1}^k \sigma_j\right]\label{eq:F1}    \\
&&~~~~+\1\left[\sum_{j=1}^k \sigma_j=0\right]\gamma(\sigma_1,\ldots,\sigma_k)       \Bigg\}\nonumber\\
&&~~= \sum_{  \sigma_1,\ldots,\sigma_k } \left[\frac{1+\rmm}{2}\right]^{(\sum_{j=1}^k\sigma_j+k)/2}\nonumber\\
&&~~~~\times \left[\frac{1-\rmm}{2}\right]^{(k-\sum_{j=1}^k\sigma_j)/2}\Bigg\{\sgn\left[\sum_{j=1}^k \sigma_j\right]\nonumber\\
&&~~~~+\1\left[\sum_{j=1}^k \sigma_j=0\right]\gamma(\sigma_1,\ldots,\sigma_k)\Bigg\}\nonumber\\
&&~~=\sum_{\ell=0}^k \binom{k}{\ell}\left[\frac{1+\rmm}{2}\right]^{\ell} \left[\frac{1-\rmm}{2}\right]^{k-\ell}\sgn[2\ell-k]~,\nonumber
\end{eqnarray}
in Equation~(\ref{eq:p-m}) for the specific choice of $\alpha\equiv\chi$ and $n=1$. This result leads  to the  function $F_\chi^1(\rmm)$ used in Equation~(\ref{eq:m}).

We are interested in how the function  $F_\chi^1(\rmm)$ behaves on the interval  $\rmm\in[-1,1]$ and how this behavior is affected by the parameter $\tanh\beta$. In order to find this out  we first rewrite $F_\chi^1(\rmm)$ as follows
\begin{eqnarray}
&&F_\chi^1(\rmm)\\
&&=\tanh(\beta)\nonumber\\
&&\times\sum_{\ell=0}^k \binom{k}{\ell}\left[\frac{1+\rmm}{2}\right]^{\ell} \left[\frac{1-\rmm}{2}\right]^{k-\ell}\nonumber\\
&&\times\left\{\1[2\ell-k>0]-\1[2\ell-k<0]\right\}.\nonumber
\end{eqnarray}
On the other hand, observe that
\begin{eqnarray}
&&\sum_{\ell=0}^k \binom{k}{\ell}\left[\frac{1+\rmm}{2}\right]^{\ell} \left[\frac{1-\rmm}{2}\right]^{k-\ell}\\
&&=\sum_{\ell=0}^k \binom{k}{\ell}\left[\frac{1+\rmm}{2}\right]^{\ell} \left[\frac{1-\rmm}{2}\right]^{k-\ell}\nonumber\\
&&\times\left\{\1[2\ell-k\geq0]+\1[2\ell-k<0]\right\}\nonumber\\
&&=1\nonumber
\end{eqnarray}
and so
\begin{eqnarray}
&&F_\chi^1(\rmm)\\
&&=\tanh(\beta)\Bigg(1-2\sum_{\ell=0}^{\tilde{k}} \binom{k}{\ell}\nonumber\\
&&~~~~~~~~~~~~~~~~~~~~~\times\left[\frac{1+\rmm}{2}\right]^{\ell} \left[\frac{1-\rmm}{2}\right]^{k-\ell}\nonumber\\
&&- \1[k\equiv0\!\!\!\!\pmod{2}]\nonumber\\
&&~~~~\times\binom{k}{k/2}\left(\left[\frac{1+\rmm}{2}\right]\left[\frac{1-\rmm}{2}\right]\right)^{k/2}\Bigg),\nonumber
\end{eqnarray}
where $\tilde{k}=\lfloor\frac{k-1}{2} \rfloor $.

Now we use the above  expression of $F_\chi^1(\rmm)$ to compute
\begin{eqnarray}
&&\frac{\rmd}{\rmd\rmm}F_\chi^1(\rmm)\\
&&=\tanh(\beta)\sum_{\ell=0}^{\tilde{k}} \binom{k}{\ell}\nonumber\\
&&\times\Bigg((k-\ell)\left[\frac{1+\rmm}{2}\right]^{\ell} \left[\frac{1-\rmm}{2}\right]^{k-\ell-1}\nonumber\\
&&-\ell\left[\frac{1+\rmm}{2}\right]^{\ell-1} \left[\frac{1-\rmm}{2}\right]^{k-\ell}\Bigg)\nonumber\\
&&+ \tanh(\beta)\frac{k}{4}\1[k\equiv0\!\!\!\!\pmod{2}]\binom{k}{k/2}\nonumber\\
&&\times\rmm\left(\left[\frac{1+\rmm}{2}\right]\left[\frac{1-\rmm}{2}\right]\right)^{k/2-1}\nonumber\\
&&=\tanh(\beta) \binom{k}{\tilde{k}+1}(\tilde{k}+1)\nonumber\\
&&\times\left[\frac{1+\rmm}{2}\right]^{\tilde{k}} \left[\frac{1-\rmm}{2}\right]^{k-\tilde{k}-1}\nonumber\\
&&+ \tanh(\beta)\frac{k}{4}\1[k\equiv0\!\!\!\!\pmod{2}]\binom{k}{k/2}\nonumber\\
&&\times\rmm\left(\left[\frac{1+\rmm}{2}\right]\left[\frac{1-\rmm}{2}\right]\right)^{k/2-1}\nonumber
\end{eqnarray}
So, using that $\tilde{k}=\lfloor\frac{k-1}{2} \rfloor $, we obtain
\begin{eqnarray}
\frac{\rmd}{\rmd\rmm}F_\chi^1(\rmm)&=&\tanh(\beta) \binom{k}{(k+1)/2}\!\!\left(\frac{k+1}{2}\right)\label{eq:dF1}\\
&&\times\left(\left[\frac{1+\rmm}{2}\right] \left[\frac{1-\rmm}{2}\right]\right)^{(k-1)/2}\nonumber
\end{eqnarray}
for $k$ odd and
\begin{eqnarray}
\frac{\rmd}{\rmd\rmm}F_\chi^1(\rmm)&=&\tanh(\beta) \binom{k}{k/2}\left(\frac{k}{4}\right)\label{eq:dF2}\\
&&\times\left(\left[\frac{1+\rmm}{2}\right] \left[\frac{1-\rmm}{2}\right]\right)^{k/2-1}\nonumber
\end{eqnarray}
for $k$ even.

Thus $\frac{\rmd}{\rmd\rmm}F_\chi^1(\rmm)>0$ for all $\rmm\in(-1,1)$ and hence $F_\chi^1(\rmm)$ is a strictly increasing function. Furthermore, the function $F_\chi^1(\rmm)$  at the point $\rmm=0$ changes its slope from $\frac{\rmd}{\rmd\rmm}F_\chi^1(\rmm)\vert_{\rmm=0}<1$ to $\frac{\rmd}{\rmd\rmm}F_\chi^1(\rmm)\vert_{\rmm=0}\geq1$ at  $$\tanh(\beta)=2^{k-1}/k\binom{k-1}{(k-1)/2}$$ for $k$ odd and $$\tanh(\beta) =2^{k-2}/\binom{k-2}{(k-2)/2}(k-1)$$ for $k$ even.

Let us now compute the second derivative of $F_\chi^1(\rmm)$. Differentiating  Equations~(\ref{eq:dF1}) and (\ref{eq:dF2}) with respect to $\rmm$ gives
\begin{eqnarray}
\frac{\rmd^2}{\rmd\rmm^2}F_\chi^1(\rmm)&=&-\rmm\tanh(\beta)\label{eq:ddF1}\\
&&\times \binom{k}{(k+1)/2}\left(\frac{k+1}{2}\right)\frac{(k-1)}{4}\nonumber\\
&&\times\left(\left[\frac{1+\rmm}{2}\right] \left[\frac{1-\rmm}{2}\right]\right)^{(k-1)/2-1}\nonumber
\end{eqnarray}
for $k$ odd and
\begin{eqnarray}
\frac{\rmd^2}{\rmd\rmm^2}F_\chi^1(\rmm)&=&-\rmm\tanh(\beta)\label{eq:ddF2}\\
&&\times \binom{k}{k/2}\left(\frac{k}{4}\right)\frac{(k-2)}{4}\nonumber\\
&&\times\left(\left[\frac{1+\rmm}{2}\right] \left[\frac{1-\rmm}{2}\right]\right)^{k/2-2}\nonumber
\end{eqnarray}
for $k$ even. We note that both are of the form $\frac{\rmd^2}{\rmd\rmm^2}F_\chi^1(\rmm)=-\rmm G(\rmm)$, where $G(\rmm)>0$ for all $\rmm\in(-1,1)$. Thus the function  $F_\chi^1(\rmm)$ is strictly convex and concave on the intervals  $(-1,0)$ and $(0,1)$ respectively.

\section*{Acknowledgment}

This work is supported by the EU FET project STAMINA (FP7-265496)  and the Leverhulme trust grant F/00 250/H.

\ifCLASSOPTIONcaptionsoff
  \newpage
\fi




\end{document}